\documentclass{webofc}
\usepackage[varg]{txfonts}   
\begin{document}
\title{Prospects for the NA60+ experiment at the CERN SPS }

\author{\firstname{Alessandro} \lastname{De Falco}\inst{1}\fnsep \thanks{\email{alessandro.de.falco@ca.infn.it}} for the NA60+ Collaboration.
}

\institute{Università/INFN Cagliari, Italy 
          }

\abstract{%
A new heavy-ion experiment on fixed target, NA60+, has been proposed at the CERN SPS for data taking in the next years. Its main goals will be focused on precision studies of thermal dimuons, heavy quark and strangeness production in Pb-Pb collisions at center-of-mass energies ranging from 5 to 17 GeV, which will provide a unique opportunity to investigate the region of the QCD phase diagram at high baryochemical potential ($\mu_B \sim 200-400$~MeV). 
The key points of the NA60+ very broad and ambitious physics program will be described.
}
\maketitle

\begin{figure}[h]
    \centering
    \includegraphics[width=0.5\textwidth]{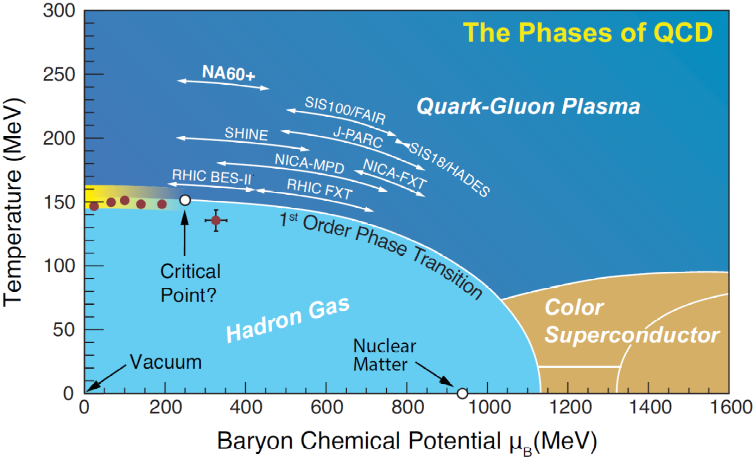}
    \caption{The QCD phase diagram (courtesy of Thomas Ullrich)}
    \label{fig:phase_diagram}       
\end{figure}

Heavy-ion collisions at low center-of-mass (CM) energies provide a unique tool to investigate the QCD phase diagram at large values of the baryochemical potential $\mu_B$. An energy scan at the SPS, with CM energies ranging from 5 to 17 GeV per nucleon, will allow the study of the phase diagram in the range $\mu_B \sim 200-400$~MeV (fig.~\ref{fig:phase_diagram}), where key information can be obtained about the presence of a critical point, the order of the phase transition at large $\mu_B$, the chiral symmetry restoration effects and the temperature at which the onset of the deconfinement takes place. 

The NA60+ experiment has been proposed~\cite{eoi} to study hard and electromagnetic probes in this $\mu_B$ region. Such measurements would address many yet uncovered points in the understanding of the QGP phase. Hard probes may bring information on the onset of deconfinement via the measurement of $J/\psi$ suppression as a function of the CM energy, while open charm states would allow to determine the transport properties of the Quark-Gluon Plasma (QGP) at large values of $\mu_B$. 
Electromagnetic probes, and in particular dileptons, would give insight on the temperature of the system via the measurement of the thermal dimuon mass spectrum, on chiral symmetry restoration effects and on the order of the phase transition~\cite{thermal,chiral}. 

\begin{figure}[t]
    \centering
    \includegraphics[width=\textwidth]{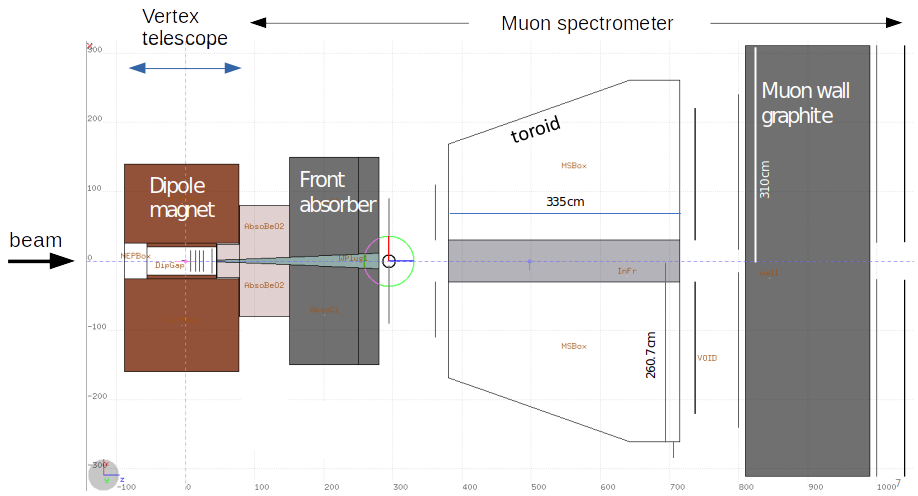}
    \caption{Layout of the proposed NA60+ apparatus in the low energy setup.}
    \label{fig:layout}       
\end{figure}

The detector concept of NA60+ inherits the design from its predecessor NA60, and is based on a muon spectrometer coupled with a vertex telescope. 
Muons are detected by a magnetic spectrometer, composed of a toroidal magnet and four tracking stations, placed after a hadron absorber. 
A new magnet, which should provide a field $B\cdot R\sim 0.2-0.5$~Tm is being designed and a small-scale prototype has been constructed. 
The tracking stations will be equipped with about 330 GEM modules, each one measuring $50 \times 110$~cm$^2$, overlapping in both horizontal and vertical coordinates by 10~cm. 
The GEM module design profits from the experience developed for the CMS muon system and the ALICE TPC upgrades.
A trigger system based on two RPC stations is placed after an additional graphite absorber. 
The width of the front absorber, as well as the position of the tracking stations will be changed depending on the beam energy, such that the detector keeps a good acceptance around mid-rapidity. The low-energy ($E_{\mathrm{beam}}/A = 20-40$~GeV) setup is reported in fig.~\ref{fig:layout}. At high energy, $E_{beam}/A=158$ GeV, the muon spectrometer will be moved on rails by 3.3~m and the absorber thickness will be increased to 4.6~m.
The presence of the hadron absorber provides the muon identification. However, it introduces a deterioration in the muon momentum resolution, due to multiple scattering and fluctuations in the energy loss. 
This loss in resolution can be recovered by matching the muons reconstructed in the muon spectrometer with the charged tracks measured in the vertex telescope (VT), both in coordinates and momentum space. The VT can also be used for multiplicity measurements and for the study of hadronic decay channels of charm and strange hadrons, as a stand-alone detector. The VT consists of a dipole magnet and a set of pixel planes for tracking. 
The choice considered for the magnet is the CERN MEP48 dipole, that provides a field $B=1.47$~T and an angular coverage of $21^0$. The tracking system will consist of a set of (5 or 10) planes made of four large area Monolithic Active Pixel Sensors (MAPS) obtained by means of the stitching technology. The pixel dimensions are of about $15 \times 15~\mu\mathrm{m}^2$. The sensor thickness is of the order of $20~\mu$m. Moreover, the mechanical supports and cooling are moved to the border. Therefore, the material budget in the sensitive area is very small, lower than $0.1\%X_0$, reducing the effect of multiple scattering in the VT. The expected spatial resolution is $5\mu$m or better, with a gain of a factor of $~2$ with respect to the hybrid technology.

A high intensity beam is required for the measurement of hard and electromagnetic probes. The experiment is foreseen to use  the H8 beam line at the CERN SPS, which can provide an intensity of $10^7$~Pb ions per 20~s spill. The interaction rate is thus expected to be one order of magnitude higher with respect to other experiments in the same $\mu_B$ range. 
With a typical run time of one month per energy, a very significant statistics can be reached for each of the proposed measurements. 

\begin{figure}[t]
    \centering
    \includegraphics[width=0.47\textwidth]{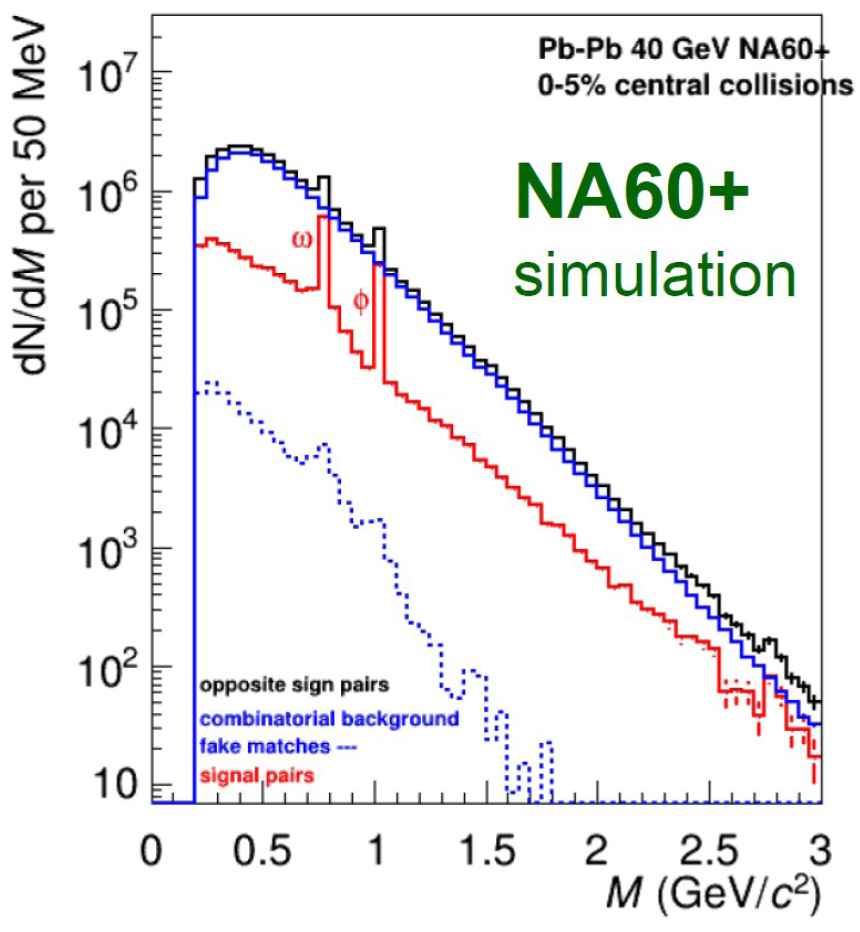}
    \includegraphics[width=0.49\textwidth]{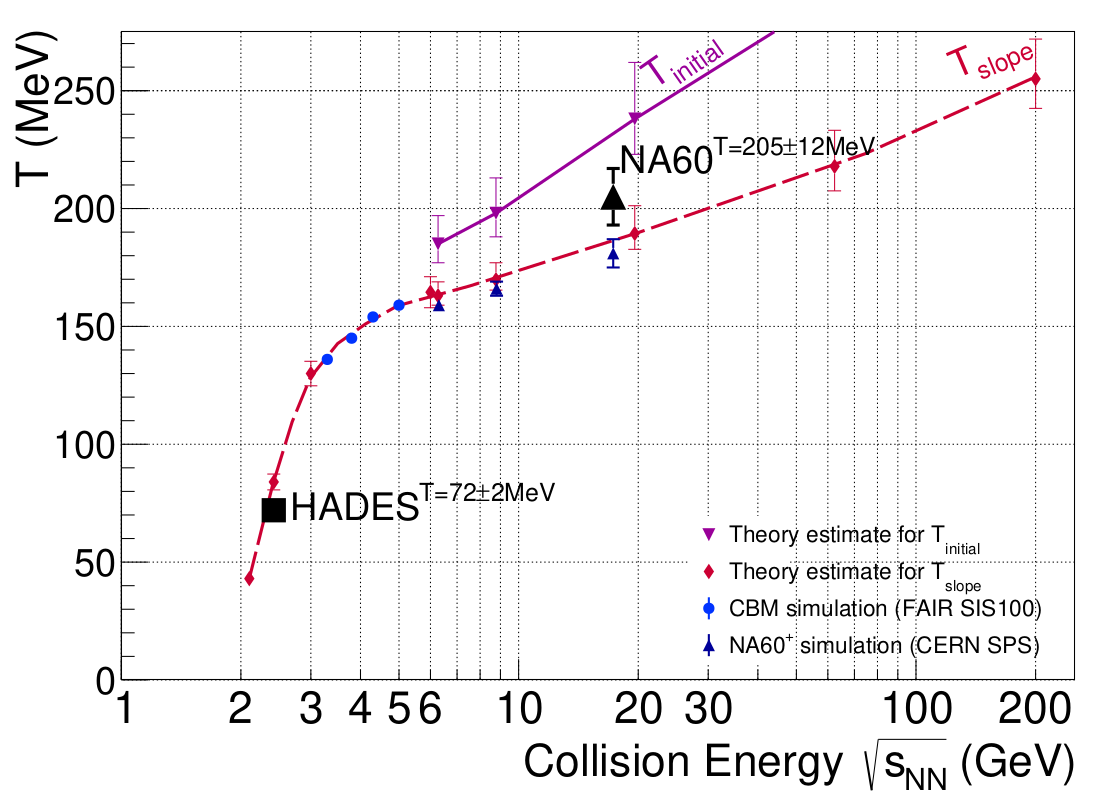}
    \caption{Left: simulated dimuon mass spectrum in central Pb-Pb collisions at $E=40$~GeV per nucleon. Right: medium temperature evolution vs $\sqrt{s_{NN}}$ in central Pb-Pb collisions using~\cite{thermal}. Measurements from NA60~\cite{NA60thermal} and HADES~\cite{hades}.}
    \label{fig:dimumass}       
\end{figure}

The physics performances of NA60+ were studied performing fast simulations of the signals with semi-analytical tracking based on the Kalman filter, while the simulation of the background was performed with FLUKA. The obtained opposite sign dimuon mass spectrum is shown in the left panel of fig.~\ref{fig:dimumass}. 
The mass resolution for $M=M_\omega$ is $\sim 7$~MeV/$c^2$, about one third of the NA60 resolution at the same mass.  
Besides the signal, composed of muon pairs, two sources of background are visible: the combinatorial background and the fake matches contribution. The former is due to uncorrelated pairs of muons mainly originated from pion and kaon decays. The latter is due to incorrect matches between muons reconstructed in the muon spectrometer and tracks reconstructed in the vertex telescope. Both contributions can be evaluated using event mixing techniques and subtracted to the opposite sign mass spectrum. The signal mass spectrum resulting after background subtraction is dominated by the hadronic cocktail for $M<1.5$~GeV/$c^2$. In this region, a precision measurement of the $\rho$ spectral function can be performed, complementing the NA60 measurement in In-In at top SPS energy with results at lower energy. 
In the region $1<M<1.5$~GeV/$c^2$, a dimuon enhancement due to the chiral mixing between the $\rho$ and its chiral partner $a_1$ via $4\pi$ states can be observed~\cite{chiral}. 
For $M>2$~GeV/$c^2$, after subtracting the Drell-Yan and open charm contributions, the mass spectrum is dominated by thermal dimuons, that can be fitted to a thermal distribution of the form $dN/dM \propto M^{-3/2} \exp(-M/T_S)$. The parameter $T_S$ represents a space-time average of the thermal temperature over the fireball evolution and can be determined with a precision of the order of 10~MeV. Measurements of $T_S$ vs $\sqrt{s_{NN}}$ at low energy ($<10$~GeV) may allow to discover a plateau in the caloric curve (fig.~\ref{fig:dimumass}, right) that would be present in case of a phase transition of the first order~\cite{thermal}. 

Charmonium suppression was extensively studied at the SPS, where the NA50 collaboration observed a suppression of $\sim 30\%$ of the $J/\psi$ meson yield in central Pb-Pb collisions at the top SPS energy ($\sqrt{s_{NN}}=17.3$~GeV) that could not be ascribed to cold nuclear matter effects alone~\cite{NA50}. The suppression is qualitatively consistent with the melting of the less bound $\chi_c$ and $\psi(2S)$ states in a deconfined medium, which would lead to a suppression of the $J/\psi$ produced from the decays of these particles. At present, no measurement is available at lower energy. NA60+ aims to extend the measurements down to $E_{\mathrm{lab}}/A=40$~GeV in order to search for the onset of the deconfinement. By correlating the measurement with the corresponding determination of the temperature obtained from thermal dimuons, one could obtain the temperature at which the $\chi_c$ and $\psi(2S)$ melt. With one month data taking at an intensity of $5 \times 10^5$ Pb ions per second, a number of $J/\psi$ in the range $1.5-20\times 10^4$, depending on the beam energy, may be reconstructed. Also cold nuclear matter effects have to be determined precisely, therefore, data taking with p-A collisions are mandatory at each CM energy. A 15-days long data taking with a beam intensity of $3\cdot 10^8$ protons/s on 7 nuclear targets is planned to measure the $J/\psi$ production cross section as a function of the mass number. 

Open charm measurements can be performed reconstructing the decays of charmed hadrons into two or three charged hadrons, using the vertex telescope as a stand-alone detector. The huge combinatorial background can be reduced by applying geometrical selections on the displaced decay vertex topology, profiting from the fact that the $c\tau$ for these decays is about $60-310\mu$m, and therefore the decay vertices are displaced by a few hundred $\mu$m from the interaction point. A high resolution on the vertex reconstruction is needed. The MAPS technology can provide a signal to background ratio $\sim 10$ times higher than the corresponding value obtainable with hybrid pixel sensors. As of today, no open charm measurement has been performed below the top SPS energy. In one month data taking, more than $3\cdot 10^6~D^0$ can be reconstructed in central collisions at $\sqrt{s_{NN}}=17.3$~GeV, allowing for a precise determination of the yield and elliptic flow as a function of $p_T$, rapidity and centrality. At $\sqrt{s_{NN}}=10.6$~GeV due to the lower production cross section, the number of reconstructed $D^0$ is expected to be lower by about an order of magnitude. Still, the measurement will be feasible with a statistical precision at the level of the percent. 

Finally, strangeness measurements in the hadronic decay channels will allow to explore the low multiplicity region for the understanding of strangeness enhancement with multiplicity. The high statistics expected to be collected will guarantee a high $p_T$-reach, the study in narrow centrality bins and an extention to multistrange hadrons of the elliptic flow measurements. 

The broad physics program of the NA60+ experiment provides a strong motivation for its realization. The goal is to start data taking in 2027, with LHC run 4.

\bibliography{bibliography}

\end{document}